\documentclass[fleqn,twoside]{article}
\usepackage{epsfig}
\usepackage{graphicx}
\usepackage[figuresright]{rotating}

\def\rQCD{{\rm QCD}}

\newcommand{\AmS}{{\protect\the\textfont2
  A\kern-.1667em\lower.5ex\hbox{M}\kern-.125emS}}

\usepackage{espcrc2}
\usepackage{amsmath}
\title{QCD soft gluon exponentiation: YFS MC Approach\thanks{Work partly supported 
by the US Department of Energy Contract  DE-FG05-91ER40627
, by NATO Grant PST.CLG.977751, and by
Polish Government grant 5P03B09320.}}

\author{B.F.L. Ward\address[MPI]{Werner-Heisenberg-Institut, 
Max-Planck-Institut feur Physik, \\ 
        Foehringer Ring 6, Muenchen, Germany}%
        \thanks{Permanent Address: Department of Physics, University of 
         Tennessee, Knoxville, TN 37996-1200, USA.}
        and
        S. Jadach\address{Henryk Niewodniczanski Institute of Nuclear Physics,
 \\
        ul. Radzikowskiego 152, 31-342 Cracow, Poland}}
\begin{document}

\begin{abstract}
We develop and prove the
theory of the QCD
extension of the YFS Monte Carlo approach to higher order
QED radiative corrections. As a corollary, a new
approach to quantum gravity by one of us (B.F.L.W.) is illustrated. 
Semi-analytical results 
and preliminary explicit Monte Carlo data are
presented for the processes
$p\bar{p}\rightarrow t\bar{t}+X$ at FNAL energies. 
We comment briefly on the implications of our results
on the CDF/D0 observations and on RHIC/LHC physics.
\centerline{UTHEP-02-0801,MPI-PhT-2002-032}
\centerline{Aug., 2002}
\vspace{1pc}
\end{abstract}

\maketitle

\section{Introduction}
The problem of soft gluon resummation is well known~\cite{sterm,cat}
and some of its many phenomenological applications are also: 
the FNAL t\=t production cross section higher order corrections
( the current situation~\cite{cdf1,d01} has 
the experimental cross section $6.2{+1.2\atop -1.1}$pb
to be compared with a theoretical prediction~\cite{wilb} of $5.1\pm 0.5$pb)
and the attendant soft gluon uncertainty in the extracted value
of $m_t = 0.1743\pm 0.0051$TeV, where~\cite{granis} $\sim 2-3$GeV
of the latter error could be due to soft gluon uncertainties;
RHIC hard scattering polarized pp scattering processes, etc.
For the LHC/TESLA/LC, the requirements on the corresponding theoretical
precisions will be even more demanding 
and the QCD soft $n(G)$ MC exponentiation which we discuss in the following
will be an important part of the necessary theory  --
YFS~\cite{yfs} exponentiated ${\cal O}(\alpha_s^2)L$
corrections realized on an event-by-event basis.

The results which we present also will allow us to investigate
from a different perspective some of the outstanding theoretical
issues in perturbative QCD, such as the treatment of phase space,
no-go theorems for the soft regime, etc. -- see ref.~\cite{app1}.

For definiteness,
we will use the process in Fig.~1 , $\bar Q(p_1) Q(q_1)\rightarrow \bar t(p_2) t(p_1) +
G_1(k_1)\cdots G_n(k_n)$, as our proto-typical process.
\begin{figure}
\begin{center}
\setlength{\unitlength}{1mm}
%%%%\begin{picture}(160,80)
\begin{picture}(80,40)
%%\put(0,0){\framebox( 65,60){ }}
%%%%\put(-2.4, -10){\makebox(0,0)[lb]{
\put(1.2, -5){\makebox(0,0)[lb]{
%\epsfig{file=epi02-fg1g.eps,width=80mm,height=30mm}
%}}
\epsfig{file=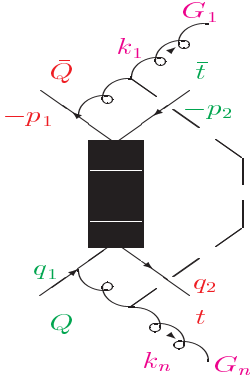,width=35mm,height=50mm}
}}
\end{picture}
\label{figproto.1}
\caption{\sf  The process $\bar{Q}
    Q      \rightarrow        \bar{t}
                 +      t    + n(G)$.
The four--momenta are indicated in the standard manner: $q_1$ 
is the
four--momentum of the incoming $Q$, $q_2$ 
is the four--momentum of the outgoing $t$, etc.,
and $Q = u,d,s,c,b,G.$}
\end{center} 
\end{figure}
\noindent
Extension 
of the methods we develop to other related
processes will be immediate -- 
one of us (B.F.L.W.) has realized
a new approach to quantum gravity as 
a by-product.
Although we shall use the older EEX  formulation of YFS
MC exponentiation as defined in ref.~\cite{sjbflw0}, the realization
of our results via the the newer CEEX formulation of YFS exponentiation
in ref.~\cite{ceex:2001} is also possible and is in progress~\cite{elsewh}.

After a brief review of the QED case, we prove our result
for the QCD case and conclude with some illstrative results.
%We organize our presentation
%as follows. In the next section, we review the application of YFS MC methods
%in the EW-QED case. In Sect. 3, we prove that we can extend this 
%application to the QCD theory and we show that this implies
%a new approach to quantum gravity. In Sect. 4 we illustrate
%the QCD extension by applying it to t\=t production at FNAL.
%Sect. 5  contains our summary remarks.
%\par

\section{Review of YFS Theory: An Abelian Gauge Theory Example}

The Abelian gauge theory example of QED has been worked-out and
realized in many applications for SLC/LEP1 and LEP2 physics in the
MC's YFS2, YFS3, BHLUMI, BHWIDE, KORALZ, ${\cal KK}$ MC, YFSWW3, YFSZZ and
KoralW by S. Jadach {\it et al.} in refs.~\cite{sjbflw0,ceex:2001,sjbflw1}. 
For example,
for the process $e^+(p_1)e^-(q_1)\rightarrow \bar{f}(p_2) f(q_2) +n(\gamma)(k_1,\cdot,k_n)$,
renormalization group improved YFS theory~\cite{bflw1987} gives {\small
\begin{eqnarray}
d\sigma_{exp}=e^{2\alpha\,Re\,B+2\alpha\,
\tilde B}\sum_{n=0}^\infty{1\over n!}\int\prod_{j=1}^n{d^3k_j\over k_j^0
}\int {d^4y\over(2\pi)^4}\nonumber\cr
\qquad e^{iy(p_1+q_1-p_2-q_2-\sum_jk_j)+D}
\bar\beta_n(k_1,\dots,k_n){d^3p_2d^3q_2\over p_2^0q_2^0}
\label{eqone}
\end{eqnarray}}
where the YFS
infrared functions $\tilde B,~B$ and the IR finite function $D$
%virtual infrared function $B$ 
are known~\cite{yfs}.
%and where we note the usual connections
%\[2\alpha\,\tilde B = \int^{k\le K_{max}}{d^3k\over k_0}\tilde S(k)\nonumber\]
%\begin{equation}D=\int d^3k{\tilde S(k)\over k^0}\left(e^{-iy\cdot k}-\theta(K_{
%max}-k)\right)\label{eqtwo}\end{equation} for the standard YFS infrared emission
%factor {\small
%\begin{equation}\tilde S(k)= {\alpha\over4\pi^2}\left[Q_fQ_{
%{\llap{\phantom f}^{\sstl(}\bar f^{\sstl^)}{}}'}\left({p_1\over p_1\cdot k}-{q_1
%\over q_1\cdot k}\right)^2+(\dots)\right]\label{eqthree}\end{equation}} 
%\noindent if $Q_f$
%is the electric charge of $f$ in units of the positron charge. 
For example, the YFS hard photon residuals $\bar\beta_i$ in (\ref{eqone}), 
$i=0,1,2$,
are given in refs.~\cite{sjbflw1} for BHLUMI 4.04 realizing the
YFS exponentiated exact ${\cal O}(\alpha)$
and LL ${\cal O}(\alpha^2)$
cross section for Bhabha scattering via a
corresponding Monte Carlo implementation
of (\ref{eqone}). 
Here, we will develop
and apply the analogous theoretical paradigm to the 
prototypical QCD higher order radiative corrections problem
for $Q\bar Q\rightarrow t\bar t +n(G)$.\par

\section{Extension to non-Abelian Gauge Theories: Proof}
 
Viewing the YFS theory as a general re-arrangement of
renormalized perturbation theory based on its IR
behavior just as the renormalization group
as a general property of renormalized perturbation theory based on its
UV(ultra-violet) behavior, for the process 
$q_i + \bar q_i \rightarrow t\bar t + n(G)$,
let the amplitude be
${\cal M}^{(n)\alpha{\bar\alpha}}_{\gamma{\bar\gamma}} 
 = \sum_{\ell}M^{(n)\alpha{\bar\alpha}}_{\gamma{\bar\gamma}\ell}$,
where $M^{(n)}_{\ell}$ is the contribution to 
${\cal M}^{(n)}$
from Feynman diagrams with ${\ell}$ virtual loops. If $S_{QCD}(k)$,
as defined in ref.~\cite{app1} is the virtual gluon IR emission factor,
we get the ``YFS representation'', by analogy with the YFS analysis,
\begin{equation}
{\cal M}^{(n)} = exp(\alpha_sB_{QCD})\sum_{j=0}^\infty{\sf m}^{(n)}_j,
\label{yfsrepv}
\end{equation}
where $\alpha_s(Q)B_{QCD}$ is the usual YFS integral~\cite{yfs,app1} 
of $S_{QCD}(k)$
and
${\sf m}^{(n)}_j$
%{1\over {j!}}\int\prod_{i=1}^j{d^4k_i\over k_i^2}
%       \beta_j(k_1,\cdots,k_j). 
%\label{irfreev}
%\end{equation}
%$\beta_i(k_1,...,k_i)$ 
do not
contain the virtual IR singularities in the product 
$S_{QCD}(k_1)\cdots S_{QCD}(k_i)$.

For the respective
real IR singularities, we proceed
as in ref.~\cite{app1} 
%to write the respective cross section as 
%\begin{eqnarray}
%  d\hat\sigma^n = {e^{2\alpha_sReB_{QCD}}\over {n !}}\int\prod_{m=1}^n
%{d^3k_m \over k^0_m}\delta(X_I-X_F) \nonumber\\
%\bar\rho^{(n)}(X) 
% {d^3p_2d^3q_2\over p^0_2 q^0_2},\nonumber\\
%\label{diff1}
%\end{eqnarray}
%where we have defined $X_I=p_1+q_1,~X_F=p_2+q_2+\sum_{i=1}^nk^0_i,~
%X \equiv (p_1,q_1,p_2,q_2,k_1,\cdots,k_n)$ and  
%\begin{equation}
%\bar\rho^{(n)}(X)=
%\sum_{color,spin} \|\sum_{j=0}^\infty{\sf m}^{(n)}_j\|^2
%\label{diff2}
%\end{equation}
so that, for the real emission factor $\tilde S_{QCD}(k)$ defined
in ref.~\cite{app1}, upon effecting the analogous YFS expansion in
$\tilde S_{QCD}(k)$ and summing on $n$, we arrive at
the ``YFS-like'' result
\begin{eqnarray}
d\hat\sigma_{\rm exp}= \sum_n d\hat\sigma^n 
         =e^{\rm SUM_{IR}(QCD)}\sum_{n=0}^\infty\int\prod_{j=1}^n{d^3
k_j\over {k^0}_j} \nonumber\\
\int{d^4y\over(2\pi)^4}
e^{iy\cdot(p_1+q_1-p_2-q_2-\sum k_j)+
D_\rQCD} \nonumber\\
*\bar\beta_n(k_1,\ldots,k_n){d^3p_2\over p_2^{\,0}}{d^3q_2\over
q_2^{\,0}}, \nonumber\\
\label{subp10}
\end{eqnarray}
where the IR finite functions ${SUM}_{IR}(QCD),~D_\rQCD$ 
%${SUM}_{IR}(QCD)=2\alpha_s ReB_{QCD}+2\alpha_s\tilde B_{QCD}(\Kmax)$,
%where the functions $\tilde B_{QCD},~D_\rQCD$ 
are defined in ref.~\cite{app1}
and the $\bar\beta_n$ are the QCD hard gluon residuals~\cite{app1}.

Each order in $\alpha_s$
must make an infrared finite contribution 
to $d\bar{\hat\sigma}_{\rm exp}\equiv 
\exp[-{\rm SUM_{IR}(QCD)}]d\hat\sigma_{\rm exp}$ 
since both $SUM_{IR}(QCD)$ and $d\hat\sigma_{\rm exp}$
are IR finite. If 
$\bar\beta^{(\ell)}_n= \tilde{\bar\beta}^{(\ell)}_n + D\bar\beta^{(\ell)}_n$,
where $\tilde{\bar\beta}^{(\ell)}_n$ is completely free of IR
divergences, then the IR finiteness just noted
%of $d\bar{\hat\sigma}_{\rm exp}$
%order-by-order in $\alpha_s$ 
allows~\cite{app1} us to conclude that
the contributions from the $\{D\bar\beta^{(\ell)}_n\}$ cancel in
$d\hat\sigma_{\rm exp}$ so that we arrive at the new result  
\begin{eqnarray}
%\begin{split}
d\hat\sigma_{\rm exp}= \sum_n d\hat\sigma^n = e^{\rm SUM_{IR}(QCD)}
         \sum_{n=0}^\infty\int\prod_{j=1}^n{d^3k_j\over k_j} \nonumber\\
\int{d^4y\over(2\pi)^4}e^{iy\cdot(p_1+q_1-p_2-q_2-\sum k_j)+
D_\rQCD} \nonumber\\
\tilde{\bar\beta}_n(k_1,\ldots,k_n){d^3p_2\over p_2^{\,0}}{d^3q_2\over
q_2^{\,0}},\nonumber\\
%\end{split}
\label{subp15}
\end{eqnarray}
where the new hard gluon residuals $\tilde{\bar\beta}_n(k_1,\ldots,k_n)=
\sum_{\ell=0}^\infty \tilde{\bar\beta}^{(\ell)}_n(k_1,\ldots,k_n)$
are completely free of IR divergences. Earlier arguments in ref.~\cite{delaney}
were insufficient to derive the analog of (\ref{subp15})~\cite{app1}.
\par

There are many consequences of (\ref{subp15}) and the theory 
which underlies it~\cite{elsewh}. We illustrate some of them
in the next section. In addition, 
one of us (B.F.L.W.) recently pointed-out~\cite{bflw2} that the YFS resummed
propagators in quantum gravity, via the formula 
\begin{eqnarray}
i\Delta'_F(k)|_{YFS-resummed} =  \frac{ie^{B''_g(k)}}{(k^2-m^2-\Sigma'_s+i\epsilon)},
\label{resum}
\end{eqnarray}
with the graviton exponent, in the deep UV, ( here, $\kappa^2=8\pi G_N$ )  
\begin{eqnarray} 
B''_g(k) = \frac{\kappa^2|k^2|}{8\pi^2}\ln\left(\frac{m^2}{m^2+|k^2|}\right),
\label{yfs1} 
\end{eqnarray}
leads to exponential damping of all propagators in the deep UV and 
thereby by to UV finite loop corrections for quantum gravity. This is then a 
new approach~\cite{bflw2} to quantum gravity.

\section{YFS Exponentiated QCD Corrections to t\= t 
Production at High Energies}

We have realized the result (\ref{subp15}) via semi-analytical
methods and via MC methods~\cite{app1}. Here we illustrate these
realizations.

For the process $p\bar p\rightarrow t\bar t+X$ at FNAL energies,
we use a semi-analytical realization of (\ref{subp15})
together with the standard formula structure function formula~\cite{app1} for
$\sigma(t\bar t+X)$, where
the DGLAP synthesization procedure presented in
ref.~\cite{sjbflw2}
%MPL{\bf A14}(1999)491  by BW and SJ 
is applied to avoid over-counting
resummation effects already included in the structure function
DGLAP evolution. In the MC realization, we employ the MC methods
of ref.~\cite{sjbflw0} to get the MC ttp1.0 as the respective event-by-event
simulation. For the semi-analytical 
analysis, we have found in ref.~\cite{sjbflw3}
%MPL{\bf A12}(1997)242
the normalization $n(G)$ effect ($r^{nls}_{exp}$ is the ratio
of the exponentiated and Born cross sections)
$r^{nls}_{exp}=1.086,~1.103,~1.110$ for $\alpha_s$ evaluated at the
scales $\sqrt s,~2m_t,~m_t$, respectively, which implies~\cite{app1} that
the corrections for ${\cal O}(\alpha_s^n,n\ge 2)$ give
$0.006-0.008$ of the NLO cross section, in agreement with ref.~\cite{cat2}.
This is an important cross check of our methods.

For the MC data from ttp1.0, which are preliminary,
show in Fig.~2 
\begin{figure}
\begin{center}
\setlength{\unitlength}{1mm}
%%%%\begin{picture}(160,80)
\begin{picture}(80,40)
%%\put(0,0){\framebox( 65,60){ }}
%%%%\put(-2.4, -10){\makebox(0,0)[lb]{
\put(-1.2, -50){\makebox(0,0)[lb]{
%\epsfig{file=epi02-fg1g.eps,width=80mm,height=30mm}
%}}
\epsfig{file=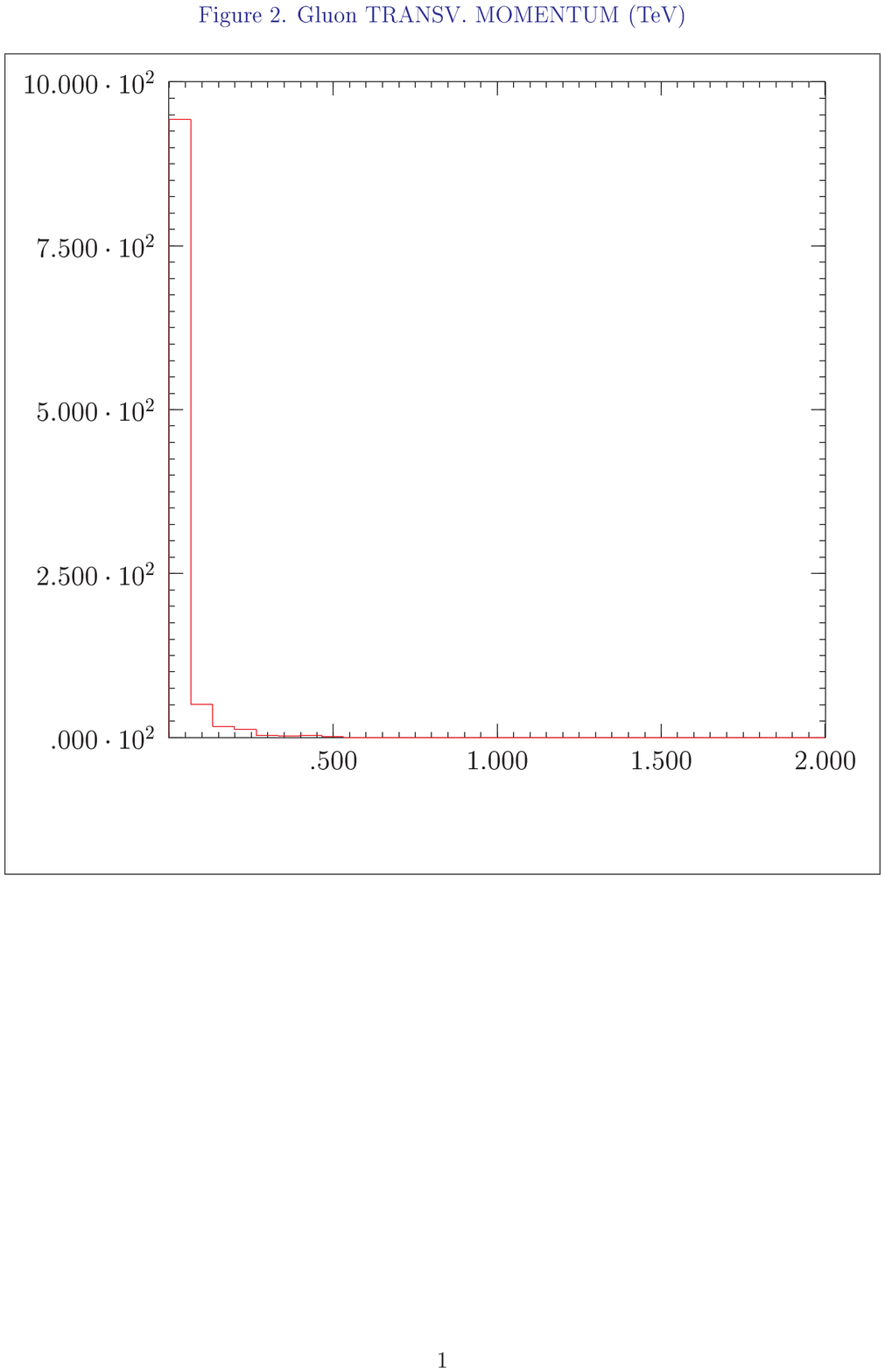,width=70mm}
}}
\end{picture}
%\caption{Fig. 2. {\small Multiple gluon $p_\perp$ distribution.}}
\label{figpperp}
\end{center} 
\end{figure}
where we see in the gluon transverse momentum distribution
that $<p_\perp>$ is indeed large at FNAL and its effect must be taken
into account in precision studies of the top quark production and decay
systematics.\par

One of us (B.F.L.W.) thanks Prof. S. Bethke for the support and kind
hospitality of the MPI, Munich, during the final stages of this work.
The authors also thank Profs.~G.~Altarelli and Prof. Wolf-Dieter Schlatter
for the support and kind hospitality 
of CERN while a part of this work was completed.
These same two authors also thank Profs.~F.~Gilman and W.~Bardeen of the
former SSCL for their kind hospitality while this work was in its development
stages.\par


\begin{thebibliography}{9}
\bibitem{sterm} G. Sterman, Nucl.Phys. {\bf B281} (1987) 310, and references
therein.
\bibitem{cat} S. Catani and L. Trentadue, Nucl.Phys. {\bf B327} (1989) 323;
{\em ibid.}{\bf B353} (1991) 183.
\bibitem{cdf1}
S. R. Blusk, in {\em Proc. ICHEP2000}, ed. C.S. Lim and T. Yamanaka,
 ( World Scientific, Singapore, 2001) p. 811.
\bibitem{d01}
D. Chakraborty, in {\em Proc. ICHEP2000}, ed. C.S. Lim and T. Yamanaka,
 ( World Scientific, Singapore, 2001) p. 814.
\bibitem{wilb} S. R. Willenbrock, in {\em Proc. RADCOR2000, Carmel, 2000}, eds.
H. Haber and S. Brodsky, eConf C000911, (2001) 379;
Rev. Mod. Phys. {\bf 72} (2000) 1141.
\bibitem{granis}P. Grannis, private communication.
\bibitem{yfs}
D.~R.~Yennie, S.~C.~Frautschi, and H.~Suura, Ann. 
Phys. {\bf 13} (1961) 379;\newline
see also K.~T.~Mahanthappa, Phys.~Rev.~{\bf 126} (1962) 329, 
for a related analysis.
\bibitem{app1} B.F.L. Ward and S. Jadach, Acta Phys. Pol. {\bf B33} (2002) 
1543, and references therein.
%\bibitem{gps} S. Jadach, B.F.L. Ward and Z. Was, Eur. Phys. J. {\bf C22}
%(2001) 423.
\bibitem{sjbflw0} S.Jadach and B.F.L. Ward, Phys.Rev. {\bf D38} (1988) 2897;
{\it ibid.} {\bf D39} (1989) 1471;{\it ibid.} {\bf D40} (1989) 3582;
Comput. Phys. Commun. {\bf 56} (1990) 351.
\bibitem{ceex:2001}
S. Jadach, B.F.L. Ward and Z. Was, Phys. Rev. D{\bf 63} (2001) 113009.
\bibitem{elsewh} S. Jadach {\it et al.}, to appear.
\bibitem{sjbflw1} 
%S.Jadach, B.F.L. Ward and Z. Was, Comput. Phys. Commun. {\bf 66} (1991) 276;
%S.Jadach and B.F.L. Ward, Phys. Lett. {\bf B274} (1992) 470;
%S. Jadach {\em et al.}, Comput. Phys. Commun. {\bf 70} (1992) 305;
%S.Jadach, B.F.L. Ward and Z. Was, Comput. Phys. Commun. {\bf 79} (1994) 503;
%S. Jadach {\em et al.}, Phys. Lett. {\bf B353} (1995) 362; {\it ibid.} 
%{\bf B384} (1996) 488; Comput. Phys. Commun. {\bf 102} (1997) 229;
%S.Jadach, W. Placzek and B.F.L. Ward, Phys. Lett. {\bf B390} (1997) 298;
%Phys. Rev. {\bf D54} (1996) 5434; Phys.Rev. D56 (1997) 6939;
%S.Jadach, M. Skrzypek and B.F.L. Ward, Phys. Rev. {\bf D55} (1997) 1206;
See, for example, S. Jadach {\em et al.}, Phys. Lett. {\bf B417} (1998) 326;
Comput. Phys. Commun. {\bf 119} (1999) 272; Phys. Rev. {\bf D61} (2000) 113010;
Phys. Rev. {\bf D65} (2002) 093010; Comput. Phys. Commun. {\bf 140} (2001) 432,
475;
%S.Jadach, B.F.L. Ward and Z. Was, Comput. Phys. Commun. {\bf 124} (2000) 233; Comput. Phys. Commun. {\bf 130} (2000) 260;
%Phys. Rev. {\bf D63} (2001) 113009; 
and references therein.
\bibitem{bflw1987} B.F.L. Ward, Phys. Rev. {\bf D36} (1987) 939.
\bibitem{delaney}
D. DeLaney {\it et al.},Phys. Rev. D{\bf 52} (1995) 108, Phys. Lett. {\bf B342} (1995) 239.
\bibitem{bflw2} B.F.L. Ward, hep-ph/0204102.
\bibitem{sjbflw2}B.F.L. ward and S. Jadach, Mod. Phys. Lett. {\bf A14} (1999)  
491.
\bibitem{sjbflw3} S. Jadach {\it et al.}, Mod. Phys. Lett. {\bf A12} (1997) 
242. 
\bibitem{cat2} S. Catani {\it et al.}, Phys. Lett. {\bf B378} (1996) 329.
\end{thebibliography}
\end{document}